\documentclass[twocolumn,showpacs,preprintnumbers,amsmath,amssymb]{revtex4}
\usepackage{graphicx}
\usepackage{dcolumn}
\usepackage{bm}

\begin{document}

\title{Coherent Betatron Oscillation and Induced Errors\\    
in the Experimental Determination of the Muon g-2 Factor}

\author{Y.N. Srivastava}
\author{A. Widom}
\affiliation{Physics Department, Northeastern University, Boston MA 02115}

\date{\today}

\begin{abstract}
Precision measurements of the anomalous magnetic moment 
of the muon depend on the correct collective mode assignment of the 
coherent betatron oscillations in the beam. Presently, there is 
a disagreement between experiment and theory for the horizontal 
betatron frequency. The discrepancy is here resolved by our 
computations of electrostatic beam focusing. The correct treatment 
of the betatron effects renders less likely the need for 
non-standard model corrections to the muon g-2.    
\end{abstract}

\maketitle

\section{Introduction}

There has been considerable interest in precision 
measurements\cite{1,2,3,4,5,6,7} 
of the muon anomalous magnetic moment factor 
\begin{equation}
g=2(1+\kappa ).
\end{equation}
The anomaly factor is measured from the chiral asymmetry  rotation 
frequency 
\begin{equation}
\omega_\kappa =\kappa \left({|e{\bf B}|\over Mc}\right).
\end{equation}
Largely due to experimental improvements in creating an extremely 
homogeneous magnetic field \begin{math} {\bf B} \end{math}, 
there have been recent improvements\cite{7} in the accuracy of 
\begin{math} \kappa \end{math} determinations. 

In addition to the anomaly frequency of Eq.(2), there are three other 
frequency scales of interest in the experiments. 
The cyclotron frequency is given by  
\begin{equation}
\omega_c=\left({c|e{\bf B}|\over {\cal E}}\right)
=\left({|e{\bf B}|\over \gamma Mc}\right),
\end{equation}
which is also determined by the homogeneous magnetic field. 
Due to electrostatic focusing capacitors, there are two more 
frequencies having to do with collective oscillations of the beam as 
a whole. These are the vertical betatron oscillations at frequency
\begin{math} \Omega_{||}  \end{math} and the horizontal betatron 
oscillations at frequency \begin{math} \Omega_\perp \end{math}. 
In the literature\cite{8,9}, the ratios 
\begin{math} (\Omega_{||}/\omega_c) \end{math}
and 
\begin{math} (\Omega_\perp/\omega_c) \end{math} 
are called, respectively, the vertical tune and horizontal tune.

Recent experiments\cite{7,9} yield a vertical tune of 
\begin{equation}
\left({\Omega_{||}\over \omega_c }\right)=\sqrt{n}
\end{equation}
with a {\em measured} effective index of   
\begin{equation}
n\approx 0.137\ .
\end{equation}
Previous theoretical considerations\cite{8,9} lead to a horizontal  
tune of  
\begin{equation}
\left({\Omega_\perp\over \omega_c }\right)_{\rm expected}
=\sqrt{1-n}\approx 0.929\ .
\end{equation}
The horizontal tune has been measured by the muon decay oscillations 
themselves. (See \cite{7} and \cite{9} wherein the notation 
\begin{math} \omega_b \end{math} is used for 
\begin{math} \Omega_\perp \end{math}.)   
\begin{equation}
\left({\Omega_\perp\over \omega_c }\right)_{\rm experiment}
\approx 0.0702\ . 
\end{equation}
The discrepancy between the previous theoretical Eq.(6) and the experimental 
Eq.(7) is evident.

One purpose of our work is to show that the previous theory is at fault.
The index \begin{math} n \end{math} describes the electric field gradient 
inside the electrostatic focusing capacitors; It is 
\begin{equation}
n=\left({\rho_0\over |{\bf B}|}\right)
\left|\left<{c\over |{\bf v}_\perp |}
{\partial E_{||}\over \partial r_{||}}\right>\right|, 
\end{equation}
where 
\begin{math} \rho_0 \approx 711.2\ {\rm cm} \end{math} 
is the radius of the cyclotron, 
\begin{math} {\bf v}=(d{\bf r}/dt)  \end{math} is the muon velocity, 
\begin{math} E_{||}=({\bf E\cdot B}/|{\bf B}|) \end{math} is the component 
of the electric field parallel to the magnetic field and  
\begin{math} r_{||}=({\bf r\cdot B}/|{\bf B}|) \end{math}. The averaging 
is over a cyclotron orbit. 

Some insight into the error of the previous theoretical Eq.(6) becomes 
evident from the fact that when the electric field vanishes, 
\begin{math}
\lim_{n\to 0}(\Omega_\perp /\omega_c)_{\rm expected}=1 
\end{math}, the horizontal tune is unity. Such ``unit horizontal tunes'' 
are in reality {\em not} collective modes at all. A charged particle in a 
uniform magnetic field (and zero electric field) moves in a circle with 
a fixed center. If the center of circular orbit is {\em not quite equal} 
to the center of the cyclotron machine (which is always the case), 
then the above corresponds to what has been called a unit horizontal 
tune.  

However, the true nature of the horizontal betatron oscillation is 
that the center of the cyclotron orbit also rotates if and only if a  
non-zero electric field is applied. A central result of this work is 
that the center of the cyclotron orbit also moves in a circle at 
a horizontal tune of  
\begin{equation}
\left({\Omega_\perp\over \omega_c }\right)
={1\over 2}\left({\Omega_{||}\over \omega_c }\right)^2 =
\left({n\over 2}\right)\approx 0.0685\ .
\end{equation}
Our theoretical Eq.(9) is in satisfactory agreement with 
the experimental Eq.(7). 

In order to understand the nature of the 
horizontal betatron tune, consider a toy ``hula hoop'' popular 
with children near the middle of the last century. Pretend there 
are particles rotating within (and relative to) the hoop 
with angular velocity \begin{math} \omega_c \end{math}. Further imagine the 
hoop rotating as a whole ``off center'' with an angular velocity 
\begin{math} \Omega_\perp \end{math}. Now remove the hoop since it is 
imaginary anyway. The remaining particle motions constitute a beam with 
a horizontal tune.

In Sec.\ II, the dynamics of the muon motion will be discussed 
starting from the Hamiltonian 
\begin{equation}
H=\sqrt{c^2({\bf p}-(e/c){\bf A})^2+c^4M^2}+e\Phi ,
\end{equation}
where the magnetostatic and electrostatic fields are given, respectively, 
by
\begin{equation}
{\bf B}=curl{\bf A}\ {\rm and}\ {\bf E}=-{\bf grad}\Phi .
\end{equation}
For simplicity of presentation, we shall employ quantum mechanics 
for describing the beam dynamics but we shall not employ (for now) the 
spin dynamics which must be treated with a Dirac Hamiltonian  
and requires somewhat long and delicate spinor algebra\cite{10,11}. 
The full quantum Dirac spinor algebra will be discussed elsewhere and 
should serve as a more accurate quantum replacement for the 
classical BMT equations\cite{12}. 

In Sec.\ III, the coherent betatron frequencies will be derived from 
the Hamiltonian Eq.(10). A physical description of the coherent betatron 
oscillations will be provided. In the concluding Sec.\ IV, the 
consequences for the \begin{math} (g-2) \end{math} 
analyses due to the corrected horizontal tune theory will be explored.

\section{Quantum Beam Dynamics}

\begin{figure}[bp]
\scalebox {0.5}{\includegraphics{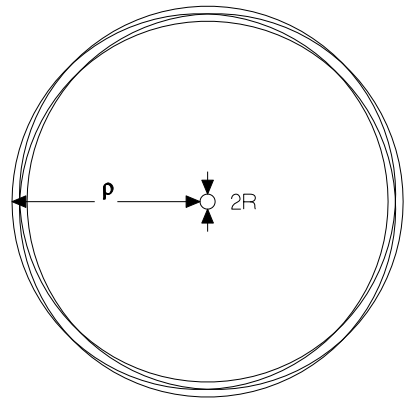}}
\caption{The electrostatic focusing field pushes on a particle in  
a cyclotron orbit of radius vector ${\bf \rho }$. The force causes 
the instantaneous center of the cyclotron orbit to rotate at angular 
velocity $\Omega_\perp $ with a radius ${\bf R}$. As schematically 
shown above, the resulting wobble in the particle orbit represents 
the horizontal coherent betatron oscillation.}
\label{fig1}
\end{figure}

The Hamiltonian in Eq.(10) may be considered to be an operator 
by employing the canonical commutation relations 
\begin{equation}
\left[p_i,r_j\right]=-i\hbar \delta_{ij}.
\end{equation}
We also consider a coordinate system in the laboratory frame 
in which the uniform magnetic field vector is along the 
\begin{math} z=r_{||}  \end{math} axis; i.e. 
\begin{equation}
{\bf A}=(A_x,A_y,0)\ \  {\rm and}\ \ {\bf B}=(0,0,B). 
\end{equation} 
From Eqs.(11)-(13) it follows that 
\begin{equation}
\left[p_x-{e\over c}A_x,p_y-{e\over c}A_y\right]
=\left({ie\hbar B\over c}\right) .
\end{equation}
The cyclotron radius vector in the horizontal plane 
\begin{equation}
{\bf \rho }=(\rho_x, \rho_y, 0)
\end{equation}
points from the cyclotron orbit center to the charged particle 
position. In mathematical terms
\begin{equation}
{\bf \rho}=\left({c\over eB^2}\right){\bf B\ \times }
\left({\bf p}-{e\over c}{\bf A}\right)
\end{equation}
The components of the cyclotron radius vector in Eq.(15) do not 
commute. From Eqs.(14)-(16)
\begin{equation}
\left[ \rho_x, \rho_y \right]=\left({i\hbar c\over eB}\right).
\end{equation}
The vector giving the position in the horizontal plane of the 
cyclotron orbit center 
\begin{equation}
{\bf R }=(R_x, R_y, 0)
\end{equation}
is defined so that 
\begin{equation}
{\bf r}={\bf r}_{||}+{\bf r}_\perp 
\ {\rm and}\  {\bf r}_\perp = {\bf \rho }+{\bf R}.
\end{equation}
The components of the cyclotron orbit center position also do not 
commute; i.e. 
\begin{equation}
\left[R_x, R_y \right]=-\left({i\hbar c\over eB}\right).
\end{equation}
We note {\em en passant} as an example of a noncommutative 
geometry\cite{13} that the center coordinates 
\begin{math} {\bf R} \end{math} of the cyclotron orbit 
live in the noncommutative geometric plane in virtue of Eq.(20).
Finally, we have the vanishing commutators: 
\begin{equation}
\left[{\bf \rho},{\bf R}\right]={\bf 0}, \ \ 
\left[{\bf \rho},z\right]=\left[{\bf R},z\right]={\bf 0},
\end{equation}
and
\begin{equation}
\left[{\bf \rho},p_z\right]=\left[{\bf R},p_z\right]={\bf 0}.
\end{equation}
The Hamiltonian of Eq.(10) now reads 
\begin{equation}
H=\sqrt{c^2p_z^2 +(eB)^2{\bf \rho }^2+c^4M^2}+
e\Phi ({\bf \rho}+{\bf R},z).
\end{equation}
To calculate the rate at which \begin{math} {\bf R} \end{math}
changes with time, one may use 
\begin{equation}
\dot{\bf R}={i\over \hbar }\left[H,{\bf R}\right]=
\left({ie\over \hbar }\right)
\left[\Phi ({\bf \rho}+{\bf R},z) ,{\bf R}\right].
\end{equation}

Thus, the center of the cyclotron orbit drifts with 
a velocity which depends only on the electromagnetic fields 
acting on the particle; i.e. 
\begin{equation}
{\bf V}=\dot{\bf R}=c\left({{\bf E\times B}\over B^2}\right).
\end{equation}
The method of deriving Eq.(25) using quantum mechanics is due to 
Schwinger\cite{14}.

In the direction parallel to the magnetic field,  
\begin{equation}
v_z={i\over \hbar }\left[H,z\right]=\left({c^2p_z\over {\cal E}}\right),
\end{equation}
where 
\begin{equation}
{\cal E}=\sqrt{c^2p_z^2 +(eB)^2{\bf \rho }^2+c^4M^2}=\gamma Mc^2.
\end{equation}
Furthermore, 
\begin{equation}
\dot{p}_z={i\over \hbar }\left[H,p_z\right]
=\left({ie\over \hbar }\right)
\left[\Phi ({\bf \rho}+{\bf R},z),p_z\right].
\end{equation}
Eqs.(27) and (28) imply 
\begin{equation}
M{d(\gamma v_z)\over dt}=eE_z.
\end{equation}

If the Hamiltonian in Eqs.(10) and (23) were treated using classical 
mechanics, then the operator equations of motion would turn into equations 
for particle orbits. There is sometimes simplicity in orbital pictures. 
Note that the classical orbital equations are recovered by  
replacing commutators with Poisson brackets:
\begin{equation}
{\rm (quantum)}\ \ {i\over \hbar}\left[a,b\right]
\ \to\ {\rm (classical)}\ \ \left\{a,b\right\}
\end{equation}
where 
\begin{equation}
\left\{a,b\right\}=\left({\partial a\over \partial {\bf p}}\right)
\cdot \left({\partial b\over \partial {\bf r}}\right)-
\left({\partial b\over \partial {\bf p}}\right)
\cdot \left({\partial a\over \partial {\bf r}}\right).
\end{equation}
In the classical orbit view wherein 
\begin{math}\dot{\gamma }=\{H,\gamma \}\end{math}, one finds 
\begin{equation}
Mc^2\left({d\gamma \over dt}\right)=e{\bf v\cdot E}.
\end{equation}

The quantum length scale for which the classical orbit picture breaks 
down can be deduced from the noncommutative geometry of the cyclotron 
center coordinate \begin{math} {\bf R}  \end{math}. From Eq.(20), 
the quantum uncertainties obey 
\begin{equation}
\Delta R_x \Delta R_y \ge (L^2/2),
\end{equation}
where 
\begin{equation}
L=\sqrt{(\hbar c/|e{\bf B}|)}\ .
\end{equation}
For the muon experiments of interest 
\begin{math} L\approx 2.0\times 10^{-6}\end{math} cm, 
which allows for a classical orbit picture of high accuracy.
Even in the classical case, the decomposition 
\begin{math} {\bf r}_\perp ={\bf \rho}+{\bf R} \end{math} 
of the position in the plane is crucial for understanding the 
horizontal betatron oscillation mode.

\section{Coherent Betatron Oscillations}

To compute the betatron vertical and horizontal tunes, one may expand
the electric field 
\begin{math}
{\bf E}({\bf \rho}+{\bf R},z)
\end{math}
in powers of the small displacements \begin{math} z \end{math} 
and \begin{math} {\bf R} \end{math}. Inside the electrostatic 
focusing capacitors  
\begin{equation}
div{\bf E}=\left({\partial E_z\over \partial z}\right)+
\left({\partial \over \partial {\bf R}}\cdot {\bf E}_\perp \right)=0.
\end{equation} 
Defining 
\begin{equation}
Q=-e\left({\partial E_z\over \partial z}\right)=
e\left({\partial \over \partial {\bf R}}\cdot {\bf E}_\perp \right),
\end{equation}
and taking only first powers of the displacement 
\begin{math} z  \end{math} in Eq.(29) yields 
\begin{equation}
M\gamma \left({d\over dt}\right)^2z= -Qz.
\end{equation}
To this accuracy, the vertical betatron oscillation frequency is 
given by 
\begin{equation}
M\gamma \Omega_{||}^2=Q 
\end{equation}
in agreement with Eqs.(4) and (8).

Expanding the electric field in Eq.(25) to linear order in 
\begin{math} {\bf R} \end{math} yields 
\begin{equation}
\left({d{\bf R}\over dt}\right)=\left({cQ\over 2eB^2}\right) 
({\bf R \times B})  
\end{equation}
where Eq.(36) has been invoked. In the \begin{math} z=0 \end{math} 
plane, the vector \begin{math} {\bf R} \end{math} rotates with 
angular velocity whose magnitude is given by 
\begin{equation}
\Omega_\perp =\left|{cQ\over 2eB}\right|,   
\end{equation}
in agreement with Eqs.(8) and (9); i.e. from Eqs.(3), (38) and (40) 
it follows that 
\begin{equation}
\Omega_{||}^2=2\Omega_\perp \omega_c\ .
\end{equation}

The horizontal oscillation is depicted in FIG.\ 1. If a particle is 
injected into the cyclotron at a somewhat skewed angle, then the center 
of the cyclotron orbit will be somewhat removed from the center of the 
cyclotron machine. The instantaneous center of the cyclotron orbit will 
itself move in a small circle at an angular velocity 
\begin{math} \Omega_\perp   \end{math}.
The small circular rotations of \begin{math} {\bf R} \end{math} 
induce a {\em wobble} in the particle's (almost) circular orbit, 
at the same frequency \begin{math} \Omega_\perp \end{math}. If a 
particle is injected into the cyclotron with a small velocity component 
\begin{math} v_z  \end{math} normal to the plane, then the height 
\begin{math} z \end{math} of the particle undergoes oscillations 
with frequency \begin{math} \Omega_{||} \end{math}. Since, in virtue 
of Eq.(36), there is but one electric field gradient parameter 
\begin{math} Q \end{math}, the two betatron frequencies are not 
independent. They are related by Eq.(41).

\section{Conclusion}

On a {\em phenomenological} basis, the chiral asymmetry of the muon decay 
is fit to experimental\cite{7} data by a function of the product form 
\begin{equation}
P_{tot}(t)=P_\kappa (t)P_{betatron}(t)P_{loss}(t),
\end{equation}
where the chiral rotation function is parameterized by 
\begin{equation}
P_\kappa (t)=e^{-t/\gamma \tau_\mu }
\left(1+A_\kappa\sin(\omega_\kappa t+\phi_\kappa )\right),
\end{equation}
the coherent horizontal betatron oscillation is parameterized by 
\begin{equation}
P_{betatron}(t)=e^{-t^2/\tau_b^2 }
\left(1+A_b \cos(\Omega_\perp t+\phi_b )\right),
\end{equation}
and the extra loss of muons from unknown physical processes  
are parameterized by
\begin{equation}
P_{loss}(t)=1+n_le^{-t/\tau_l}.
\end{equation}
The agreement between the experimental best fit\cite{9} 
\begin{equation}
\left({\Omega_\perp \over \omega_\kappa }\right)_{experiment}
\approx 2.05, 
\end{equation}
and our theoretical prediction based on Eqs.(8) and (9)
\begin{equation}
\left({\Omega_\perp \over \omega_\kappa }\right)_{theory}
\approx 2.01,  
\end{equation}
gives us confidence that we have properly identified the source 
of horizontal betatron oscillations. It is something of a mystery 
to us why the experiment was designed so that 
\begin{math} \Omega_\perp  \end{math} is so very close to being 
a ``second harmonic'' of the anomaly frequency 
\begin{math} \omega_\kappa  \end{math}. 

If one wishes to go beyond the phenomenological view of how the 
electromagnetic field 
\begin{math} F_{\mu \nu }=\partial_\mu A_\nu -\partial_\nu A_\mu \end{math} 
affects the coupling \begin{math} g=2(1+\kappa ) \end{math}, 
one must employ the Dirac-Schwinger equation 
\begin{equation}
\left\{-i\hbar \gamma^\mu d_\mu +Mc-
\left({\hbar e\kappa \over 4Mc^2}\right)
\sigma^{\mu \nu }F_{\mu \nu }\right\}\psi =0,
\end{equation}
where 
\begin{math} d_\mu =(\partial_\mu -(ie/\hbar c)A_\mu ) \end{math}.
The interaction between the electromagnetic field and the anomalous 
magnetic moment is thereby 
\begin{equation}
{\kappa \over 2}\sigma_{\mu \nu }F^{\mu \nu}=
\kappa ({\bf B\cdot \sigma }-i\gamma_5 {\bf E\cdot \sigma }).
\end{equation}

The electric field coupling of Eq.(49) via 
\begin{math} {\bf E}({\bf \rho }+{\bf R},z) \end{math}
contains those degrees of freedom 
\begin{math} (p_z,z,\rho_x,\rho_y) \end{math} which enter into 
the kinetic energy \begin{math}{\cal E}=Mc^2\gamma \end{math}
in Eq.(27) and also the cyclotron center 
coordinates \begin{math} (R_x,R_y) \end{math} which rotate as in 
Eq.(39). 

The experimentally observed coupling of the chiral rotation 
frequency \begin{math} \omega_\kappa \end{math} 
to the horizontal betatron frequency  \begin{math} \Omega_\perp  \end{math} 
is mainly due to the variation in \begin{math} {\bf R} \end{math}. 
This coupling to \begin{math} {\bf R} \end{math} 
dominates the electric field corrections to 
\begin{math} \omega_\kappa \end{math} due to 
\begin{math} {\bf \rho } \end{math}. 

In assessing the electric field corrections to 
\begin{math} \omega_\kappa \end{math}\cite{7},
due to the variations in the radius 
\begin{math} ({\bf r_\perp }={\bf \rho }+{\bf R }) \end{math} 
of the beam orbit, the effect was doubly counted: 
(i) The time variations in \begin{math} {\bf r}_\perp  \end{math} 
were first counted in the function \begin{math} P_{betatron}(t) \end{math}. 
(ii) The time variations in \begin{math} {\bf r}_\perp  \end{math} 
were counted yet again as a BMT electric field correction. 
Leaving out the over counting of the electric field corrections makes 
the experimental need for non-standard model corrections to the  
muon \begin{math} (g-2) \end{math} less compelling.

\begin{acknowledgments}
We wish to acknowledge useful correspondence with R.M. Carey, 
F.J.M. Farley, J.P. Miller and B.L. Roberts.
\end{acknowledgments}


\begin{thebibliography}{99}
\bibitem{1} G. Charpak et. al. {\it Il Nuovo Cimento} {\bf 37}, 1241 (1965).
\bibitem{2} J. Bailey et. al. {\it Phys. Lett.} {\bf B28}, 287 (1968). 
\bibitem{3} J. Bailey et. al. {\it Il Nuovo Cimento} {\bf A9}, 369 (1972). 
\bibitem{4} J. Bailey et. al. {\it Nuc. Phys.} {\bf B150}, 1 (1979). 
\bibitem{5} R.M. Carey et. al. {\it Phys. Rev. Lett.} {\bf 82}, 1632 (1999).
\bibitem{6} H.N. Brown et. al. {\it Phys. Rev.} {\bf D62}, 091101 (2000).
\bibitem{7} H.N. Brown et. al. {\it Phys. Rev. Lett.} {\bf 86}, 2227 (2001).
\bibitem{8} W. Flegel and F, Krienen {\it Nuc. Inst. Meth.} {\bf 113}, 
549 (1973).
\bibitem{9} O. Rind, et. al., {\it SLAC Workshop on $e^+e^-$ Physics at 
Intermediate Energies}, April 2001, {\bf arXiv:hep-ex}/0106101 (2001).
\bibitem{10} A. Widom and Y.N. Srivastava, 
{\bf arXiv:hep-ph}/9612290 (1996).
\bibitem{11} Y. Srivastava, E. Sassaroli and A. Widom, in {\em Les 
Rescontres de Physique de la Vall\'ee d'Aoste, Results and Perspectives in 
Particle Physics}, Ed. M. Greco, March 1996, World Scientific, Singapore 
(1997).
\bibitem{12} V. Bargmann, L. Michel and V.L. Telegdi, 
{\it Phys. Rev. Lett.} {\bf 2}, 435 (1959). 
\bibitem{13} A. Connes, {\it Noncommutative Geometry}, Academic Press, 
New York (1994).
\bibitem{14} J. Schwinger, {\it Quantum Mechanics: Symbolisim of Atomic 
Measurements}, Ed. B. Englert, Chapter 8, Springer, Berlin (2001).




\end{thebibliography}
\end{document}